\documentclass[aps,prc,superscriptaddress,twoside,twocolumn,nofootinbib,10pt,%
showpacs,floatfix]{revtex4-1}

\usepackage{amsmath,amssymb}
\usepackage{graphicx,bm}
\usepackage{slashed}
\usepackage{epstopdf}
\usepackage{ulem} 
\usepackage[usenames]{color}
\usepackage{subfigure}
\usepackage{float}

\newcommand{\Slash}[1]{{\ooalign{\hfil/\hfil\crcr$#1$}}}

\allowdisplaybreaks

\begin{document}

\title{Mass degeneracy of the heavy-light mesons with chiral partner structure in the half-skyrmion phase}

\author{Daiki Suenaga}
\email{suenaga@hken.phys.nagoya-u.ac.jp}
\affiliation{Department of Physics,  Nagoya University, Nagoya, 464-8602, Japan}

\author{Bing-Ran He}
\email{he@hken.phys.nagoya-u.ac.jp}
\affiliation{Department of Physics,  Nagoya University, Nagoya, 464-8602, Japan}

\author{Yong-Liang Ma}
\email{yongliangma@jlu.edu.cn}
\affiliation{Center for Theoretical Physics and College of Physics, Jilin University, Changchun, 130012, China}

\author{Masayasu Harada}
\email{harada@hken.phys.nagoya-u.ac.jp}
\affiliation{Department of Physics,  Nagoya University, Nagoya, 464-8602, Japan}

\date{\today}

\newcommand\sect[1]{\emph{#1}---}
\begin{abstract}
We explore the mass splitting of the heavy-light mesons with chiral partner structure in nuclear matter. In our calculation, we employed the heavy hadron chiral perturbation theory with chiral partner structure and the nuclear matter is constructed by putting skyrmions from the standard Skyrme model onto the face-centered cubic crystal and regarding the skyrmion matter as nuclear matter. We find that, although the masses of the heavy-light mesons with chiral partner structure are splitted in the matter-free space and skyrmion phase, they are degenerated in the half-skyrmion phase in which the chiral symmetry is restored globally. This observation suggests that the magnitude of the mass splitting of the heavy-light mesons with chiral partner structure can be used as a probe of the phase structure of the nuclear matter.
\end{abstract}
\maketitle

Although the nuclear matter properties are difficult to access, it is a crucial and an interesting object to study them in both particle and nuclear physics because they are critically concerned with such issues as the equation of state (EoS) relevant to the compact-star matter and the chiral
symmetry breaking/restoration in dense matter( see., e.g., Ref.~\cite{Lee:2011tz} and references therein).

Among all the approaches to the nuclear matter, skyrmion crystal is such one in which the nuclear matter properties are studied by putting skyrmions onto the crystal structure and regarding the skyrmion matter as baryonic matter~\cite{Klebanov:1985qi}(see also Ref.~\cite{PV09} and references therein). By changing the crystal size, the density effect enters. For example, in the face-centered cubic (FCC) crystal~\cite{Kugler:1988mu,Kugler:1989uc} adopted in this paper, $\rho = 4/(2L)^3$ with $\rho$ and $L$ being the nuclear matter density and crystal size, respectively. The advantage of the skyrmion crystal approach to nuclear matter is that both the nuclear matter and medium modified hadron properties can be treated in a unified way~\cite{Lee:2003aq}.

In the skyrmion crystal approach, when we reduce the crystal size, or, equivalently, increase the nuclear matter density, the nuclear matter undergoes a phase transition from skyrmion phase to half-skyrmion phase in which there is a skyrmion configuration with a half baryon number at each crystal vertex~\cite{Wuest:1987rc}. And people found that, when the skyrmions are put onto the FCC crystal at low density, in the half-skyrmion phase at high density, the crystal vertices at which half-baryons are concentrated form a cubic crystal~\cite{Kugler:1988mu,Kugler:1989uc}. The order parameter which charactorizes this phase transition is the space average of the quark-antiquark condensate $\langle \bar{q}q\rangle$ which vanishes in the half-skyrmion phase. Note that although the space average of the quark-antiquark condensate vanishes in the half-skyrmion phase, chiral symmetry is still locally broken since the pion decay constant in the baryonic matter $f_\pi^{\ast}$ which charactorizes the chiral symmetry breaking does not vanish~\cite{Ma:2013ooa} and the quark-antiquark condensate is locally non-zero~\cite{Ma:Prepare}. At this moment, properties of the half-skyrmion phase are not well-known except those pointed above.

Since in Skyrme model, there exists a well-known spin-isospin correlation, in Ref.~\cite{Suenaga2014}, we proposed to study the medium modified mass spectra of the ground states of the heavy-light mesons to probe the structure of the spin-isospin correlation in the nuclear matter constructed from the FCC crystal skyrmion matter and chiral density wave nuclear matter.
It was shown that the spin-isospin correlation generates a mixing among the heavy-light mesons carrying different spins and isospins, and that the structure of the mixing reflects the pattern of the correlation, i.e. the remaining symmetry.
Furthermore, it was found that the magnitude of the mass modification provides information of the strength of the correlation.

In this work, we focus on the the mass spectra of the heavy-light mesons in the half-skyrmion phase. Since the half-skyrmion phase is characterized by the vanishing of the space average of the quark-antiquark condensate, or ``chiral condensate", it is convenient to use the heavy-light meson fields with the chiral partner structure. Here, we regard the charmed heavy-light mesons with spin-parity quantum numbers $J^P=(0^-,1^-)$ and $J^P=(0^+,1^+)$ as chiral partners to each other~\cite{heavy-partner} which should be degenerated when the chiral symmetry is restored~\cite{Friman:2004jh,Sasaki:2014asa}. In our calculation, we take the heavy quark limit of the heavy-light mesons, so that, in the rest frame of the nuclear matter, they are at rest. In such a case, only the space averages of the relevant fields affect to the mass spectra, which enables us to study the global structure of the medium.

What we find in this paper is the following: By using the Lagrangian written up to the terms including one derivative, due to the symmetry structure of the FCC crystal and the arrangement of the nearest two skyrmions to yield the strongest attractive interaction, the mass splitting between the chiral partners is proportional to $\langle \phi_0\rangle \propto \langle \bar{q}q\rangle$ so that they are degenerate in the half-skyrmion phase. In this sense, the medium modified magnitude of the mass spitting of chiral partners can be regarded as a probe of the phase structure of the skyrmion matter.


We write the heavy-light meson doublets in the chiral basis, which, at quark level, are schematically written as $\mathcal{H}_{L,R} \sim Q\bar{q}_{L,R}$. In the present work, we only couple the heavy-light meson fields to the pion field $U(x)$. Under chiral transformation, the pion field $U(x)$ and heavy-light meson fields $\mathcal{H}_{L,R}$ transform as
\begin{eqnarray}
U & \to & g_L U g_R^{\dagger}\ ,\quad \mathcal{H}_{L,R} \to \mathcal{H}_{L,R}g_{L,R}^{\dagger}. \label{eq:ChiTrans}
\end{eqnarray}
Then, up to the one-derivative terms, the effective Lagrangian which preserves the heavy quark symmetry and $SU(2)_L \times SU(2)_R$ chiral symmetry can be constructed as
\begin{eqnarray}
{\cal L} & = &  {\rm tr}\left[\mathcal{H}_L(iv\cdot\partial)\bar{\mathcal{H}}_L] + {\rm tr}[\mathcal{H}_R(iv\cdot\partial)\bar{\mathcal{H}}_R\right] \nonumber\\
&& +\frac{1}{2}\Delta_M{\rm tr}\left[\mathcal{H}_LU\bar{\mathcal{H}}_R+\mathcal{H}_RU^{\dagger}\bar{\mathcal{H}}_L\right] \nonumber\\
&& -i\frac{g_{A_1}}{2}{\rm tr}\left[\mathcal{H}_R\gamma_5\gamma^{\mu}\partial_{\mu}U^{\dagger}\bar{\mathcal{H}}_L-\mathcal{H}_L\gamma_5\gamma^{\mu}\partial_{\mu}U\bar{\mathcal{H}}_R\right] \nonumber\\
&& -i\frac{g_{A_2}}{2}{\rm tr}\left[\mathcal{H}_L\gamma_5\Slash{\partial}UU^{\dagger}\bar{\mathcal{H}}_L-\mathcal{H}_R\gamma_5\Slash{\partial}U^{\dagger}U\bar{\mathcal{H}}_R\right] \ , \label{pionlagrangian}
\end{eqnarray}
where $v^{\mu}$ is the velocity of heave-light mesons, and $g_{A_1}, g_{A_2}$ are real parameters.

The chiral fields $\mathcal{H}_{L,R}$ relate to the heavy-light meson doublets $H$ and $G$ with quantum numbers $(0^-,1^-)$ and $(0^+,1^+)$, respectively, through
\begin{eqnarray}
\mathcal{H}_R = \frac{1}{\sqrt{2}}\left[G + iH\gamma_5\right] \ ,\ \ \mathcal{H}_L = \frac{1}{\sqrt{2}}\left[G - i H\gamma_5\right]\ . \label{eq:HLRGH}
\end{eqnarray}
In terms of the physical states, the $H$ and $G$ doublets are expressed as
\begin{eqnarray}
H &=& \frac{1+\Slash{v}}{2}\left[i\gamma_5D + \Slash{D}^\ast \right] , \quad G = \frac{1+\Slash{v}}{2}\left[D_0^\ast - i\Slash{D}_1'\gamma_5 \right] \ . \label{eq:HGphys}
\end{eqnarray}
Then, we can rewrite the effective Lagrangian~\eqref{pionlagrangian} in terms of $H$ and $G$ fields as
\begin{widetext}
\begin{eqnarray}
{\cal L} & = & {\rm tr}\left[G\left(iv\cdot\partial\right)\bar{G}-H\left(iv\cdot\partial\right)\bar{H}\right]\nonumber\\
& &{} +\frac{1}{4}\Delta_M {\rm tr}\left[G\left(U+U^{\dagger}\right)\bar{G}+H\left(U+U^{\dagger}\right)\bar{H} - iG\left(U-U^{\dagger}\right)\gamma_5\bar{H}+iH\left(U-U^{\dagger}\right)\gamma_5\bar{G}\right]\nonumber\\
& &{} - \frac{ig_{A_1}}{4}{\rm tr}\left[G\gamma_5\left(\Slash{\partial}U^{\dagger}-\Slash{\partial}U\right)\bar{G}-H\gamma_5\left(\Slash{\partial}U^{\dagger}-\Slash{\partial}U\right)\bar{H} + iG\left(\Slash{\partial}U^{\dagger}+\Slash{\partial}U\right)\bar{H}-iH\left(\Slash{\partial}U^{\dagger}+\Slash{\partial}U\right)\bar{G} \right] \nonumber\\
& &{} - \frac{ig_{A_2}}{4}{\rm tr}\left[G\gamma_5\left(\Slash{\partial}UU^{\dagger}-\Slash{\partial}U^{\dagger}U\right)\bar{G} + H\gamma_5\left(\Slash{\partial}UU^{\dagger}-\Slash{\partial}U^{\dagger}U\right)\bar{H}\right] \nonumber\\
& &{} - \frac{ig_{A_2}}{4}{\rm tr}\left[-iG\left(\Slash{\partial}UU^{\dagger}+\Slash{\partial}U^{\dagger}U\right)\bar{H} + iH\left(\Slash{\partial}UU^{\dagger}+\Slash{\partial}U^{\dagger}U\right)\bar{G}\right]\ . \label{pionlagrangian2}
\end{eqnarray}
\end{widetext}
From this Lagrangian, we see that, in the matter-free space, the $\Delta_M$ term accounts for the mass difference between $G$ and $H$ doublets whose masses in the vacuum are estimated by the spin-averaged ones as
\begin{eqnarray}
M_G & = & \frac{m_{D_0^\ast}+3m_{D_1}}{4}, \quad M_H  =  \frac{m_D+3m_{D^\ast}}{4},
\end{eqnarray}
which, by using the empirical values, lead to $M_G=2.40$~GeV, $M_H = 1.97$~GeV. Then, $\Delta_M$ can be fixed as
\begin{eqnarray}
\Delta_M= M_G - M_H = 430~\mbox{MeV}.
\end{eqnarray}
Although using the present data we can fix the combination $g_{A_1}+g_{A_2}$ through decay  $D^{*}\to D\pi$~\cite{Harada:2012km}, we do not want to specify its value here since it will be shown later that neither $g_{A_1}$ term nor $g_{A_2}$ term modifies the spectrum.

To explore the density dependence of the mass difference between $G$ and $H$ doublets, we use the skyrmion crystal approach~\cite{Ma:2013ooa} by putting skyrmions onto the FCC crystal and regarding the skyrmion matter as baryonic matter. In such an approach, the matter affects on the heavy-light meson in the medium through functions of the space-averaged classical configurations of the light meson fields. For example, for a quantity $X$, its matter effect enters through
\begin{eqnarray}
\langle X\rangle=\frac{1}{(2L)^3}\int_0^{2L} d^3xX\ ,
\label{eq:defvevX}
\end{eqnarray}
where $2L$ denotes the crystal size. In this work, we will construct the skyrmion matter by using the standard Skyrme model and, following Ref.~\cite{Lee:2003aq}, take $f_\pi = 93~$MeV and $e = 4.75$ which are the empirical values to reproduce the pion dynamics.

In our calculation, the medium modified $G$ and $H$ doublet masses are defined by the poles of the medium modified two-point functions of $G$ and $H$ doublets at the rest frame $v^{\mu}=(1,\vec{0})$ with zero residual momentum limit for the external line. Equivalently, this means that one just needs to replace the light meson fields in the Lagrangian~(\ref{pionlagrangian2}) with the space-averaged ones. Since the space-averaged light meson fields depend on the matter density, the density dependence of the heavy-light meson masses can thus be obtained. In terms of the space-averaged quantities, the effective Lagrangian~\eqref{pionlagrangian2} which is responsible for the medium modified heavy-light meson masses can be written as
\begin{widetext}
\begin{eqnarray}
{\cal L}_{\rm eff} & = & {\rm tr}\left[G\left(iv\cdot\partial\right)\bar{G}-H\left(iv\cdot\partial\right)\bar{H}\right] +\frac{1}{4}\Delta_M {\rm tr}\left[G\left\langle U+U^{\dagger}\right\rangle\bar{G}+H\left\langle U+U^{\dagger}\right\rangle\bar{H}\right]\nonumber\\
& &{} -\frac{ig_{A_1}}{4}{\rm tr}\left[G\gamma_5\left\langle\Slash{\partial}U^{\dagger}-\Slash{\partial}U\right\rangle\bar{G}-H\gamma_5\left\langle\Slash{\partial}U^{\dagger}-\Slash{\partial}U\right\rangle\bar{H}\right] \nonumber\\
& & {} - \frac{ig_{A_2}}{4}{\rm tr}\left[G\gamma_5\left\langle\Slash{\partial}UU^{\dagger}-\Slash{\partial}U^{\dagger}U\right\rangle\bar{G} +H\gamma_5\left\langle\Slash{\partial}UU^{\dagger}-\Slash{\partial}U^{\dagger}U\right\rangle\bar{H}\right] \ . \label{EffectiveLagrangian}
\end{eqnarray}
\end{widetext}
Note that in this expression, the terms mixing $G$ and $H$ doublets disappear at the rest frame since the $G$ and $H$ doublets have the opposite parity, and in the strong processes and also in our skyrmion crystal approach, the parity should be preserved.

For convenience, we next symbolically write the pion field $U$ as
\begin{eqnarray}
U= \phi_0+i\tau^a\phi^{a} , \mbox{~with~} a = 1,2,3,
\label{eq:paraU}
\end{eqnarray}
with constraint $(\phi_0)^2 + (\phi^{a})^2 = 1$. The parametrization \eqref{eq:paraU} tells us $\langle \phi_0 \rangle \propto \langle \bar{q}q \rangle$. Due to the parity conservation, we can conclude
\begin{eqnarray}
\langle\phi^{a}\rangle = 0\ .
\end{eqnarray}

In terms of $\phi^\alpha, (\alpha = 0,1,2,3)$ we have
\begin{eqnarray}
{\cal L}_{\rm eff} & = &  {\rm tr}\left[G\left(iv\cdot\partial\right)\bar{G} - H\left(iv\cdot\partial\right)\bar{H}\right]\nonumber\\
& &{} + \frac{1}{2}\Delta_M\langle\phi_0\rangle {\rm tr}\left[G\bar{G}+H\bar{H}\right]\nonumber\\
& &{} - \frac{g_{A1}}{2}\sum_i\left\langle\partial_i\phi^{i}\right\rangle{\rm tr}\left[G\gamma_5\gamma^i\tau^i\bar{G}-H\gamma_5\gamma^i\tau^i\bar{H}\right] \nonumber\\
& &{} + \frac{g_{A2}}{2}\sum_i\langle T_i\rangle{\rm tr}\left[G\gamma_5\gamma^i\tau^i\bar{G} - H\gamma_5\gamma^i\tau^i\bar{H}\right]  \ ,
\end{eqnarray}
where
\begin{eqnarray}
\langle T_i\rangle \equiv \left\langle\phi_0\partial_i\phi^{i}-\partial_i\phi_0\phi^{i}\right\rangle\ .
\end{eqnarray}
Therefore the heavy-light meson masses are modified by quantities $\langle \phi_0\rangle, \langle\partial_i\phi^{i}\rangle$ and $\langle T_i\rangle$.

Before making the numerical simulation, we first analyze some properties of $\langle\partial_i\phi^{i}\rangle$ and $\langle T_i\rangle$ based on the symmetry structure of FCC crystal and the arrangement of the nearest two skyrmions to yield the strongest attractive interaction. In the crystal, due to the periodical structure, we can expand $\phi^\alpha$ as~\footnote{Here, different from Refs.~\cite{Kugler:1989uc,Lee:2003eg}, we make the Fourier expansion of the fields $\phi^\alpha, (\alpha = 0,1,2,3)$ which have the same structures that their corresponding unnormalized ones $\bar{\phi}^\alpha$. The Fourier coefficients $\alpha$ and $\beta$ are constrained by $(\phi_0)^2 + (\phi^{a})^2 = 1$.}
\begin{eqnarray}
\phi_0 & = & \sum_{a,b,c}\beta_{abc} \cos(a\pi x/L)
\cos(b\pi y/L) \cos(c\pi z /L),\nonumber\\
\phi^1 & = & \sum_{h,k,l}\alpha_{hkl} \sin(h\pi x/L)
\cos(k\pi y/L) \cos(l\pi z/L),
\nonumber\\
\phi^2 & = & \sum_{h,k,l}\alpha_{hkl} \cos(l\pi x/L)
\sin(h\pi y/L) \cos(k\pi z/L),
\nonumber\\
\phi^3 & = & \sum_{h,k,l}\alpha_{hkl} \cos(k\pi x/L)
\cos(l\pi y/L) \sin(h\pi z/L). \label{eq:FEphi}
\end{eqnarray}
Due to the FCC structure and the arrangement of the nearest two skyrmions to yield the strongest attractive interaction, the modes appearing in above equation are restricted as follows~\cite{Lee:2003aq}:
\renewcommand{\theenumi}{\arabic{enumi}}
\renewcommand{\labelenumi}{(F\theenumi)}
\begin{enumerate}


\item $a,{}b,{}c$ are all even numbers or odd numbers and if $h$ is even, then $k,{}l$ are restricted to odd numbers while  if $h$ is odd, then $k,{}l$ are restricted to even numbers. \label{M3}
\end{enumerate}

By using definition \eqref{eq:defvevX}, from expansion \eqref{eq:FEphi}, one can easily check
\begin{eqnarray}
\langle \partial_i \phi^i \rangle & = & 0 .
\end{eqnarray}
We next consider $\langle T_i \rangle$. From the Fourier expansion \eqref{eq:FEphi} one can get
\begin{eqnarray}
\langle \phi_0 \partial_1 \phi^{1} \rangle & \sim & \sum_{a,b,c}\sum_{h,k,l} h \delta_{ah}\delta_{bk}\delta_{cl} = 0,
\end{eqnarray}
where the restriction (F\ref{M3}) has been used. A similar argument leads to
\begin{eqnarray}
\langle \phi_0 \partial_2 \phi^{2} \rangle & = & \langle \phi_0 \partial_3 \phi^{3} \rangle = 0 , \nonumber\\
\langle \partial_1 \phi_0 \phi^{1} \rangle & = & \langle \partial_2 \phi_0 \phi^{2} \rangle = \langle \partial_3 \phi_0 \phi^{3} \rangle = 0.
\end{eqnarray}
We then finally conclude
\begin{eqnarray}
\langle T_i \rangle & = & 0.
\end{eqnarray}

Then, the effective Lagrangian~(\ref{EffectiveLagrangian}) is reduced to
\begin{eqnarray}
{\cal L}_{\rm eff} & = & {\rm tr}\left[G\left(iv\cdot\partial\right)\bar{G}-H\left(iv\cdot\partial\right)\bar{H}\right]\nonumber\\
& &{} +\frac{1}{2}\Delta_M\left\langle\phi_0\right\rangle {\rm tr}\left[G\bar{G}+H\bar{H}\right] \ ,\label{eq:ga1ga20}
\end{eqnarray}
where the density affects the heavy-light meson masses through $\left\langle\phi_0\right\rangle$, or, equivalently, $\langle \bar{q}q \rangle$. This conclusion agrees with that obtained in Ref.~\cite{Harada:2003kt} in the matter-free and zero temperature space.

\begin{figure}[htbp]
\includegraphics[scale=0.3]{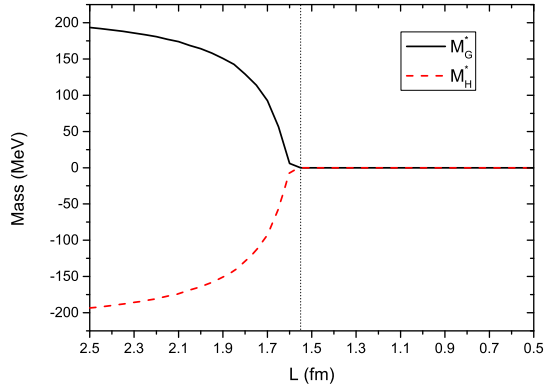}
\caption{(Color online) Medium modified heavy-light meson mass splitting as a function of the crystal size $L$. The vertical line indicates the critical density of the skyrmion-half-skyrmion phase transtion.}
 \label{eigen1}
\end{figure}

We plot in Fig.~\ref{eigen1} the crystal size $L$ dependence of the heavy-light meson masses. From this figure we see that, in the low density region (large $L$), there are two splitted lines with the upper black line denotes the modified $G$ doublet mass $M_G^*$ and the lower red-dashed line denotes the modified $H$ doublet mass $M_H^*$ and with the increase of the matter density, the mass splitting becomes smaller upto a critical density $n_{1/2}$ at which the skyrmion phase goes to half-skyrmion phase. In the half-skyrmion phase, due to the vanishing of $ \langle\phi_0\rangle \propto \langle \bar{q}q\rangle = 0$, the $H$ and $G$ doublets become degenerated.

We want to say that, since in the present case there is no spin-isospin correlation inside $G$ doublet or $H$ doublet in Lagrangian~\eqref{eq:ga1ga20}, as pointed in Ref.~\cite{Suenaga2014}, there is no mixing inside these doublets and no mass splitting between the mesons inside them.

In this paper, by using the mass splitting of the heavy-light mesons with chiral partner structure, we investigated the symmetry pattern of the half-skyrmion phase through the $G$ and $H$ doublets. To write down the effective Lagrangian, we used the chiral basis for light quarks inside the heavy-light mesons which is convenient for constructing the effective Lagrangian with chiral partner structure. In our calculation, we only consider the pseudoscalar meson, pion, effect on the heavy-light mesons and the Skyrme model is only constructed from pion, i.e., the standard Skyrme model.

Our result explicitly reveals that, in the half-skyrmion phase, due to the vanishing of the space averaged quark-antiquark condensate, the $H$ and $G$ doublets which are regarded as chiral partners to each other, have the same masses. In this sense, the medium modified mass splitting of $H$ and $G$ doublets can be used as a probe of the existence of the half-skyrmion phase. It was recently found that in a temperature system the degeneracy of chiral partners is due to the chiral symmetry restoration~\cite{Sasaki:2014asa}. However, in our present cold dense system, although the chiral partners are degenerated in the half-skyrmion phase, chiral symmetry is not restored in this phase.

In the present exploration, we only included the pion effect. The effects of the heavier resonances, such as $\rho, \omega, \sigma$ and so on which are essential for nuclear matter properties~\cite{Ma:2013ooa,Ma:2013ela}, on the heavy-light meson spectrum will be reported elsewhere.

\;
\;


We are grateful to M. Rho for valuable
comments. The work of Y.-L.~M. was supported in part by National Science Foundation of China (NSFC) under
Grant No.~10905060 and No.~11475071 and the Seeds Funding of Jilin University. M.~H. was supported in part by the JSPS Grant-in-Aid for Scientific Research (S) No.~22224003 and (c) No.~24540266.


\end {document}